\documentclass[10pt,aps,longbibliography,prd,superscriptaddress,twocolumn,amsmath,amssymb,floatfix,
nofootinbib %this makes footnotes appear below the page, and not in the bibliography like the revtex class does longbibliography
]{revtex4-1}   
           
\usepackage[dvips]{graphics}
\usepackage{bm}
\usepackage{float}

\usepackage{epsfig}
\usepackage{enumerate}
\usepackage{subfigure}
\usepackage{color}
\usepackage{graphicx} 
\usepackage[dvipsnames]{xcolor} 
\usepackage{graphicx}
\usepackage[colorlinks,citecolor=blue,linkcolor=blue,urlcolor=blue,plainpages=false,pdfpagelabels]{hyperref}    

\DeclareMathOperator{\sech}{sech}

\usepackage{tcolorbox}
\usepackage{bbold}

\newcommand{\beq}{\begin{equation}}
\newcommand{\eeq}{\end{equation}}
\newcommand{\bea}{\begin{eqnarray}}
\newcommand{\eea}{\end{eqnarray}}

\usepackage{multibib}
\newcites{supp}{Supplementary References}

\begin{document} 
\title{Photoinduced metallic Volkov-Pankratov states in semi-Dirac material} 

\author{SK Firoz Islam} 
\affiliation{Department of Physics, Jamia Millia Islamia, New Delhi-110025, INDIA}

\begin{abstract}
We study the interfacial modes across the interface between the two regions of a semi-Dirac type material, which are illuminated by the left and right circularly polarized light, respectively. We show that a smooth boundary between the two regions give rise to a momentum dependent Pöscll-Teller potential well which hosts the metallic and gapless Volkov-Pankratov states. We also note that incident electron can fully pass over the interfacial well without any reflection only at certain light parameters-known as {\it Ramsauer-Townsend} effect. Moreover, we also observe that such modes can even exist across the interface between the irradiated and non-irradiated regions under certain condition.
\end{abstract}  
\maketitle
%-------------------------------------------
\emph{Introduction}.
%-------------------------------------------
Since the discovery of photoinduced topological phase transition\cite{PhysRevB.79.081406} the research interest in the field of non-equilibrium electronic systems has grown up sharply in theory\cite{PhysRevB.85.125425,PhysRevB.89.235416,PhysRevLett.110.026603,PhysRevB.89.121401,PhysRevB.99.205429,cayssol2013floquet,PhysRevB.98.235424,PhysRevB.108.155308,PhysRevB.102.201105} as well as in experimental front \cite{lindner2011floquet,peng2016experimental,zhang2014anomalous,wang2013observation}. An electronic system can be set to in a non-equilibrium state by applying an external time dependent periodic perturbation in the form of light or irradiation. Such external perturbation can break the time reversal symmetry and leads to a non-trivial phase- known as Floquet topological phase transition\cite{PhysRevB.79.081406}. Apart from this, light can also be used to manipulate the spin and valley degrees of freedom in different types of electronic systems \cite{PhysRevB.85.205428,PhysRevB.108.155408,PhysRevB.104.L081411,PhysRevLett.116.016802,PhysRevB.84.195408,PhysRevB.98.075422,PhysRevB.90.125438}. It has been also noted that the thermoelectric properties of some Dirac-like materials can be significantly improved by exposing them to light \cite{tahir2014tunable,PhysRevB.91.115311,PhysRevB.102.045417}. A series of photo induced Weyl nodes modulation have also been reported \cite{PhysRevLett.117.087402,PhysRevB.94.235137,PhysRevB.97.155152,PhysRevB.94.041409,PhysRevB.96.041205}. Additionally, shining light on the normal region of a Josephson junction can also induce a $0-\pi$ phase transition in its transport signatures \cite{PhysRevB.94.165436,PhysRevB.95.201115}. Moreover, suitably controlling the phase of the incident light in different regions of the electronic system, a pair of interfacial chiral modes can also be engineered in three-dimensional threefold topological semimetal \cite{PhysRevB.100.165302}. More recently, higher order topological insulators have also been predicted in driven quantum systems \cite{PhysRevB.100.085138}.

The gapless one-dimensional ($1$D) edge modes or two-dimensional ($2$D) surface modes are the hallmark of topological materials \cite{RevModPhys.82.3045} with insulating bulk. Prior to the topological insulator, in early 80's, Volkov and Pankratov showed in a series of works \cite{volkov,pankratov} that one-dimensional massless metallic modes and a number of massive modes can coexist across the interface between the two regions of a semiconductor with inverted mass term. The massless $1$D modes are insensitive to the boundary details which can be identified as topological in nature whereas the massive modes are exclusively originated from the smoothness of the interface and depends on the width of the interface over which the mass term changes sign. These massive states are non-topological in nature and dubbed as Volkov Pankratov (VP) states \cite{PhysRevB.96.201302}.  The abrupt sign change of the mass term does not result into VP states, rather yields only massless  topological modes. The massless $1$D chiral topological mode has been predicted in a number of works with the inverted mass term in $2$D Dirac-type materials \cite{PhysRevB.89.085429,PhysRevLett.101.087204,PhysRevB.91.241404}, where the mass term changes sign abruptly. Unidirectional chiral interfacial electromagnetic wave has been also predicted in a broken time reversal symmetric Weyl semimetal \cite{PhysRevB.92.115310}. For the smooth interface, the VP states were obtained in photonic graphene with inverted mass term \cite{PhysRevLett.100.013904}. In recent times, a series of theoretical works investigating the VP states in different types of newly emerged electronic systems have been carried out \cite{PhysRevB.96.201302,PhysRevResearch.2.023146,PhysRevB.100.195412,PhysRevB.102.155311,PhysRevResearch.2.023373}. An experimental observation of the VP states was also reported recently \cite{PhysRevB.96.195104}.  

A graphene-like honeycomb lattice but with two different hopping parameters, $t_1$ and $t_2$, can give rise to deformed Dirac-like spectrum. It has been shown \cite{banerjee2009tight,dietl2008new} that for a special case $t_2$ = $2t_1$, a semi-Dirac type band dispersion emerges in the vicinity of $\Gamma$ point.  This kind of lattice could be realized into the three-unit-cell slab of $VO_2$  bounded by a thin insulating $TiO_2$\cite{banerjee2009tight,dietl2008new}. By using an effective low-energy Hamiltonian, a series of theoretical studies were carried out as collective excitations \cite{PhysRevB.93.085145}, nonlinear magnetotransport \cite{PhysRevB.108.035203}, irradiation effects on electronic bulk band structure \cite{PhysRevB.97.035422,PhysRevB.91.205445} including Floquet topological phase transition \cite{PhysRevB.94.081103}. Recently, the lattice model has been used to investigate the integer quantum Hall effect \cite{PhysRevB.102.085416} and topological phase in presence of a Haldane mass term \cite{mondal2021topological} too.

 In this work, we first revisit the band structure modulation of irradiated semi-Dirac material. Then we consider an interface between two regions of the semi-Dirac material that are exposed to the left and right-handed circularly polarized light, respectively. We show that contrary to the inverted mass term, the VP states emerge with inverted velocity term induced by the irradiation. Additionally, we also reveal that even the interface between non-irradiated and irradiated region can host the VP states under certain condition.

%-------------------------------------------
\emph{Model Hamiltonian and anisotropic energy spectrum}.
%-------------------------------------------
 Let's consider that the system semi-Dirac (SD) material is in the $x-y$ plane. The low energy model Hamiltonian that describe the SD material can be taken as \cite{banerjee2009tight,dietl2008new,PhysRevB.94.081103} $H=v_Fk_x\sigma_x+(\delta-\alpha k_y^2)\sigma_y$, where ${\bf \sigma}\equiv\{\sigma_x,\sigma_y\}$ are Pauli matrices in orbital space, ${\bf k}\equiv (k_x,k_y)$ are the $2$D momentum operators, $v_F$ is fermi velocity along the $k_x$ direction, $\delta$ is the gap parameter and $\alpha=(2m)^{-1}$ is a constant which inversely related to the effective mass of the fermion. The corresponding energy spectrum can be written as $E_{\zeta}=\lambda\sqrt{(\delta-\alpha k_y^2)^2+v_F^2k_x^2}$ where $\zeta\equiv\{\lambda,k_x,k_y\}$ with $\lambda=\pm$ being the band index. The energy is strongly anisotropic which disperses linearly along the $k_x$-direction and quadratically along the $k_y$ direction with the two band touching points at $\pm\sqrt{\delta/\alpha}$, bearing the significance of its name semi-Dirac (SD) material. Note that $\delta<0$ opens a gap and convert the SD into a trivial insulator. However, we shall restrict our discussion into metallic regime $\delta>0$.%%%%%%%%%%%%%%%%%%%%%%%%%%% FIGURE %%%%%%%%%%%%%%%%%%%%%%%   

%-------------------------------------------
\emph{Band structure modulation by external irradiation.}
%-------------------------------------------
Let us now consider that the SD material is exposed to an external time-dependent periodic perturbation in the form of irradiation, propagating along the $z$-direction i.e., normal to the plane of the SD material. The light field can be described by a vector potential $\mathbf{A}(t)=[A_x\sin(\Omega t+\phi), A_y \sin(\Omega t)]$, where $(A_x,A_y)$ are the field amplitudes of it's $(x,y)$ component, $\Omega$ is the frequency of the irradiation, and $\phi$ is the phase. The effects of irradiation can be inserted into the non-perturbed Hamiltonian via canonical momentum as $\mathbf{k}\rightarrow \mathbf{k}+e\mathbf{A}(t)$, where $e<0$ is the electron charge. 
To solve the periodically perturbed Hamiltonian, we can use the Floquet theory \cite{RevModPhys.89.011004}, which states that such a perturbed Hamiltonian exhibits a complete set of orthonormal solutions of the form $\psi(t)=\phi(t)e^{-i\mathcal{E} t}$ with $\phi(t)=\phi(t+T)$ being the corresponding Floquet states, $T$  is the period of the field and %cite[Eckart]%.  
 $\mathcal{E}$ denotes the Floquet quasi-energy. The time-dependent Schr{\"o}dinger  equation corresponding to the Floquet states yields the Floquet eigenvalue equation as $H_{F}\phi(t)=\mathcal{E}\phi(t)$ with $H_{F} = H(t)-i (\partial/\partial t)$. The Floquet states can be further expressed as $\phi(t)=\sum_{n}\phi{_n}(t)e^{in\Omega t}$ where $n$ is the Fourier component or Floquet side-band index. By diagonalizing the Floquet Hamiltonian in the basis of Floquet side bands `$n$', one can obtain the Floquet quasi-energy spectrum. On the other hand, an  effective Hamiltonian can be obtained in the high-frequency limit following the Floquet-Magnus expansion in power of $1/\Omega$ as $H_{eff}\simeq H+H_F^{(0)}+H_F^{(1)}+\ldots$, where the zeroth-order correction is $H_F^{(0)}=\delta_0\sigma_y$ with $\delta_0=\alpha(eA_y)^2$, and first order correction is given by $H_{F}^{(1)}=[H_{-},H_{+}]/\Omega$ where
\begin{equation}
 H_{m}=\frac{1}{T}\int_0^{T}V(t)e^{-im\Omega t}dt
\end{equation}
with $m=\pm 1$. Here, $V(t) ={\bf g(t)\cdot\sigma}$ where $g_x(t)=ev_FA_x\sin(\Omega t+\phi)$ and $g_y(t)=2e\alpha k_yA_y\sin(\Omega t)-e^2A_y^2\cos(2\Omega t)$. %The high frequency limit is described by the fact that all the energy scale related to the electronic system are much smaller than the energy of irradiation i.e., $(v_F k_F, \alpha k_F^2)<<{\Omega}$ where $k_F$ is fermi vector corresponds to standard 2D carrier density.
The first-order correction term is obtained as $H_{F}^{(1)}=v_{\Omega} k_y\sigma_z$ which is just a momentum dependent gap parameter. Here the Floquet induced velocity term $v_{\Omega}=(4e^2\alpha v_FA^2/\Omega)\sin\phi$ for $A_x=A_y=A$. Note that the sign of $v_{\Omega}$ changes when the light polarization is changed from left to right circularly polarized i.e., when phase $\phi=\pi/2\rightarrow\phi=-\pi/2$. In this context, we should mention that such velocity inverted term was considered for the surface of two dissimilar $3$D topological insulators in Refs.~[\onlinecite{PhysRevB.85.245402,PhysRevLett.107.166805}], but the interface was considered to be abrupt instead of smooth for which it would not give VP states except a single gapless unidirectional topological mode. The effective Hamiltonian can now be diagonalized to obtain the Floquet energy spectrum as 
\begin{equation}
 \mathcal{E}_{\zeta}=\lambda\sqrt{(\delta'-\alpha k_y^2)^2+v_F^2k_x^2+v_{\Omega}^2k_y^2}
\end{equation}
with $\delta'=\delta-\delta_0$ . Here gaps open at $k_y^c=\pm\sqrt{\delta'/\alpha}$, that is $2|v_\Omega k_y^c|$.

\begin{figure}[t]
 \includegraphics[width=9.2cm,height=6cm]{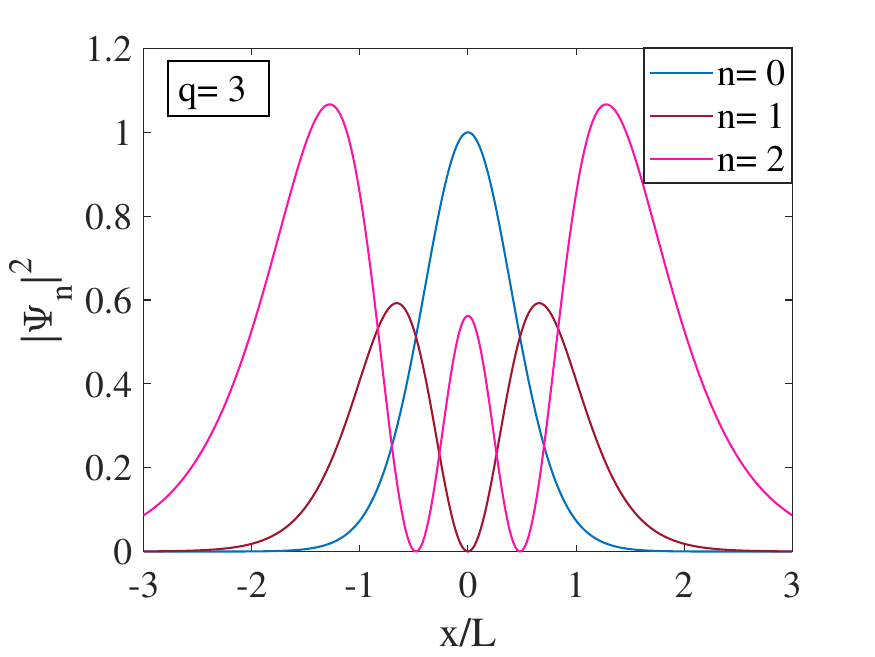}
\caption{Plots of first few VP states wave function including the zeroth topological mode, across the smooth interface. The wave functions are normalized by the factor $A_n$.} 
\label{localization}
\end{figure}

%-------------------------------------------
\emph{Interfacial modes: Volkov-Pankratov states}.
%-------------------------------------------
Now we consider that the two neighbouring regions of the SD material are exposed to the left and right circularly polarized light. The phase $\phi$ can be tuned externally by the light source, and it is taken to be $\pi/2$ and $-\pi/2$ in two regions, respectively. As a result the light induced velocity term on both regions are of opposite sign which can be modelled as $v_\Omega(x)=v_\Omega\tanh(x/L)$ where $L$ is the width of the boundary over which the sign change occurs. The interfacial modes across the boundary can be obtained by solving energy eigen value equation $H_{eff}\Psi_\zeta(x,y)=E\Psi_\zeta(x,y)$ where $\Psi_\zeta(x,y)=[\psi_1,\psi_2]^{T}e^{i k_{y}y}$. This equation can be squared on both sides to arrive at the following eigen value equation 
\begin{equation}
\Big[-\frac{\partial^2}{\partial \eta^2}+V(\eta)\Big]\psi(\eta)=\frac{L^2}{v_F^2}\varepsilon^2\psi(\eta)
\end{equation}
 where  $\eta=x/L$, $\varepsilon=\sqrt{E^2-(\delta'-\alpha k_y^2)-v_\Omega^2k_y^2}$, $\psi=i\psi_1+\psi_2$ and $ V(\eta)=-q\left(q+1\right){\sech^2(\eta)}$ with $q=Lv_\Omega k_y/v_F$. It is interesting to note that the interfacial potential $V(\eta)$ acts as a quantum-well (known as Pöschl-Teller potential well) for the incoming electrons with $q>0$ and $q<-1$. On the other hand, the potential acts as a barrier for $-1<q<0$. It is noteworthy to mention that such Pöscl-Teller type potential well was also found at the interface between the two regions of oppositely signed mass terms in an isotropic Dirac-type two band semimetal \cite{volkov,PhysRevLett.100.013904}. However, the key difference with the present case is that the depth of the well does not depend on the momentum. The bound states for the potential-well corresponds to $q>0$ can be immediately obtained as \cite{landau2013quantum}
 \begin{eqnarray}\label{volkov1}
 E_0(q)&=&\lambda(\delta'-\alpha' q^2)\\
 E_{n}(q)&=&\lambda\sqrt{E_0^2(q)+n\frac{v_F^2}{L^2}\left(2q-n\right)},\label{volkov2}
 \end{eqnarray}
 where $n=1,2,3..$ $\alpha'=\alpha v_F^2/(v_\Omega L)^2$ and $E_0^2(q)+2nq (v_F/L)^2>(nv_F/L)^2$. The corresponding eigen states can be also written as \cite{Morse1,freitas2023generalization}
 \begin{equation}
 \Psi_{n}(x)= A_n \cosh^{-(q-n)}\left(\frac{x}{L}\right)P_n^{(1,1)}\left[\tanh\left(\frac{x}{L}\right)\right]
 \end{equation}
 where $A_n$ is the $n$ dependent normalization constant, and $P_n^{a,b}$ is the Jacoby polynomial. Here the term $\cosh^{-(q-n)}(x)$ describes the localization of each VP modes, which causes each bound states die out rapidly with the distance from $x=0$. The length scale over which each bound state modes penetrate inside the bulk (localization length) are not same as shown in the Fig.~(\ref{localization}). It can be seen that the with increasing $n$, the wave function becomes more spread out with respect to the centre of the interface.  %%%%%%%%%%%%%% Rosen Morse %%%%%%%%%%%%%%%%%%%%%%%%%%%%%%%%%
%%%%%%%%%%%%%%%%%%%%%%%%%%%%%%%%%%%%%%%%%%%%%%%%%%%%%%%%

\begin{figure}[t]
 \includegraphics[width=9.2cm,height=6cm]{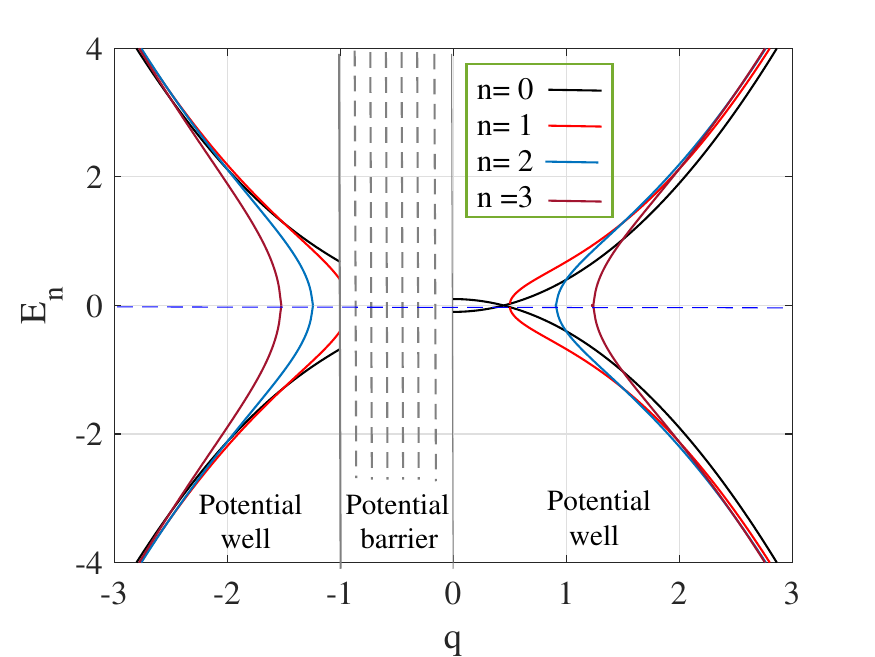}
\caption{The Volkov-Pankratov states including the zeroth mode are plotted with $q$. The energy is normalized by $E_{F}=v_Fk_F$. It can be seen that the VP states are strongly sensitive to the sign of $q$.   Typically one can take $L\sim (v_F/v_\Omega)k_F^{-1}\sim 10^{-6}$m  for a standard $2$D electron density $n_e\sim 10^{15}m^{-2}$  (that gives $k_F=\sqrt{2\pi n_e}\sim10^7m^{-1}$) and $v_F=10v_\Omega$.}
\label{volkov}
\end{figure}

  Whereas for $q<-1$, bound states are obtained as
 \begin{equation}
 E_{n}(q)=\lambda\sqrt{(\delta'-\alpha' q^2)^2-(n+1)\frac{v_F^2}{L^2}\left(2|q|-n-1\right)}
 \end{equation}
 where $n=0,1,2..$ and $(\delta'-\alpha' q^2)^2>(n+1)\left(2|q|-n-1\right)(v_F/L)^2$. Note that for $q>0$ the zeroth mode does not depend on the boundary width $L$ whereas $n>0$ modes are sensitive to the boundary details indicating that these modes are exclusively arising due to the smoothness of the boundary which can be identified as VP states. On the contrary, for $q<-1$ all the modes including $n=0$ are sensitive to the boundary width $L$,  and all of these modes are VP states. Unlike the case of graphene where the zeroth mode is unidirectional chiral \cite{PhysRevLett.100.013904}, here the zeroth mode is not chiral as can be also seen from the Fig.~(\ref{volkov}). Also, contrary to the graphene where the VP states are all gapped, these states in SD material are gapless metallic for $q>0$. On the other hand, for $q<-1$ the first few VP modes are gapped and rests are gapless, as seen in the Fig.~(\ref{volkov}). Note that the band touching point of each VP states remain protected, only moves along the $q$. It is also interesting to note that for $q>0$, all the VP states intersect with the zeroth mode at $q_c=n/2$, as can be seen from the Eq.~(\ref{volkov2}). We also note that in this region the VP states intersect with each other i.e., $n$-th and $(n+j)$-th VP states intersect at $q^{*}=n+j/2$ where $E_{n+j}-E_n=0$. Similar intersection can be also seen for $q<0$ but with slightly shifted in $q$.
%%%%%%%%%%%%%% Rosen Morse %%%%%%%%%%%%%%%%%%%%%%%%%%%%%%%%%
%%%%%%%%%%%%%%%%%%%%%%%%%%%%%%%%%%%%%%%%%%%%%%%%%%%%%%%%

Now we comment that the zeroth mode can be even directly obtained by considering an abrupt interface which can be modelled as $v_\Omega(x)=v_\Omega[2\Theta(x)-1]$ where $\Theta(x)$ is the heavyside step function. Such consideration will lead an eigen value equation
 \begin{equation}
  \left[-\frac{\partial^2}{\partial x^2}+V(x)\right]\psi(x,y)=\frac{\varepsilon^2}{v_F^2}\psi(x,y)
 \end{equation}
 where $V(x)=-2 (q/L)\delta(x)$ is a $1$D delta potential well for $q>0$ , and hosts a pair of bound states $E_0=\lambda (\delta'-\alpha' q^2)$ which are exactly the same as zeroth mode in Eq.~(\ref{volkov1}) confirming the topological nature of this mode. For $q<0$, $V(x)$ acts as a usual delta potential barrier, hence no bound states are seen there.
 
 {\it Stability of VP states} The VP states are sensitive to the boundary width $L$ and non-topological in nature hence one could expect that these states might be sensitive to a small deformation of the interfacial Pöchl-Teller potential well.  We can quickly check the fate of these states in the deformed potential well which is in fact the more realistic situation. In order to do so, we model the smooth sign change of the velocity term by using a deformed tangent hyperbolic function as $v_\Omega(x)=v_\Omega\tanh_\nu(x)$. The deformed hyperbolic function can be defined as: $\cosh_\nu(x)=(e^x+\nu e^{-x})/2$ and $\sinh_\nu(x)=(e^x-\nu e^{-x})/2$ where $0<\nu<1$ measures the degree of deformation. Such deformed hyperbolic function was introduced by Arai \cite{ARAI199163} which satisfies the relation $1-\tanh^2_\nu(x)=\nu \sech^2_\nu(x)$, $\cosh_\nu^2(x)-\sinh_\nu^2(x)=\nu$ and $\frac{d}{dx}\tanh_\nu(x)=\nu \sech^2(x)$. Under such consideration, the interfacial potential well becomes deformed Pöschl-Teller well for which the bound state solution \cite{deformed_soln} was found to be unaffected by the degree of deformation $\nu$. Hence, we can confirm that the VP states are robust to the deformation of the interfacial potential well. 
  
\begin{figure}[hbt!]
 \includegraphics[width=8.5cm,height=6cm]{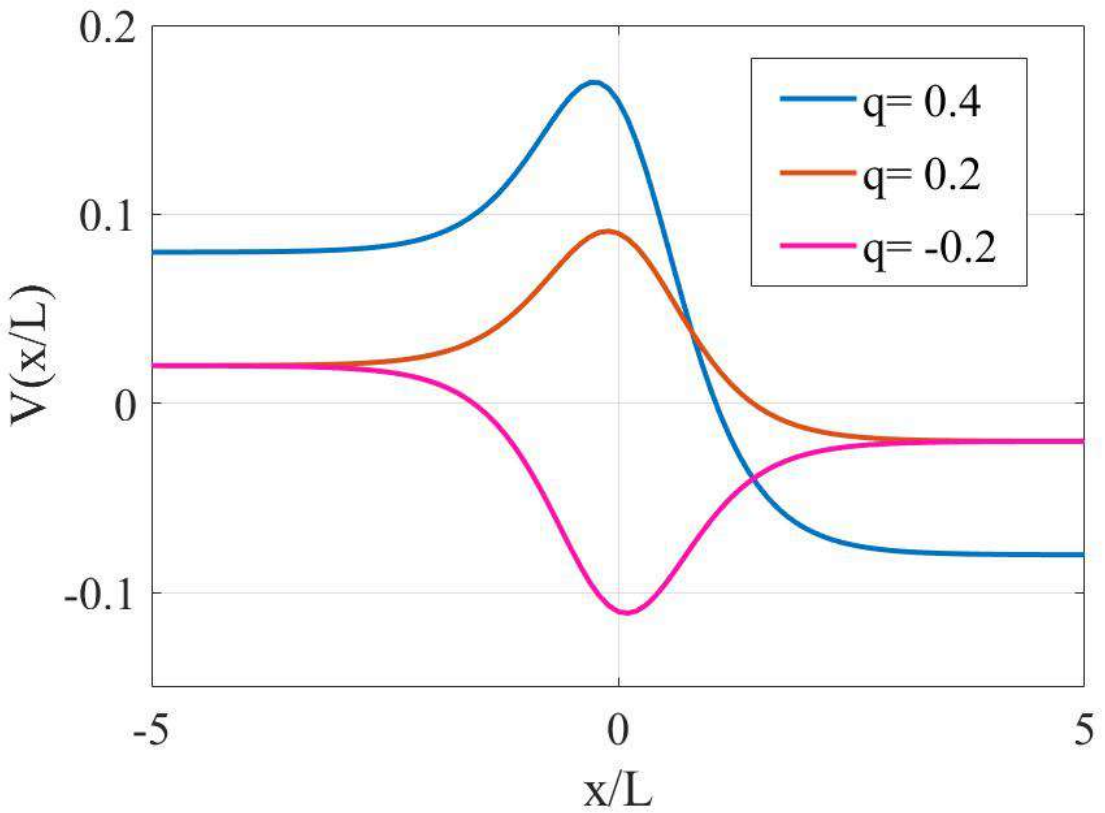}
\caption{The emergence of the potential well with varying $q$. It is clearly seen that the potential steps change to well with sigh change of $q$.} 
\label{rosen}
\end{figure}

 %------------------------------------------
\emph{Interface between irradiated and non-irradiated regions}
%-------------------------------------------
Now we consider a scenario where only one region is exposed to the irradiation. In this case, the irradiation induced velocity term can be modelled as $v_\Omega(x)=(v_\Omega/2)[1-\tanh(x/L)]$ which smoothly vanishes at the interface instead of changing sign. Following the same approach as in the previous case, we arrive at
\begin{equation}
\Big[-\frac{\partial^2}{\partial \eta^2}+V(\eta)\Big]\psi(\eta)=\frac{L^2}{v_F^2}\varepsilon^2\psi(\eta)
\end{equation}
  where
  \begin{equation}
  V(\eta)=-\frac{q^2}{2}\tanh(\eta)-\frac{q}{2}\left(\frac{q}{2}-1\right)\sech^2(\eta).
  \end{equation}
 This potential exhibits a minima across the interface for $q<0$, that appears to be an asymmetric quantum well (known as Rosen-Morse potential well \cite{Morse1}) whereas it acts as a barrier for $q>0$. The emergence of such photo tuneable well is shown in the Fig.~(\ref{rosen}). 
 %%%%%%%%%%%%%% Rosen Morse %%%%%%%%%%%%%%%%%%%%%%%%%%%%%%%%%
%%%%%%%%%%%%%%%%%%%%%%%%%%%%%%%%%%%%%%%%%%%%%%%%%%%%%%
 It is noteworthy to mention here that such quantum well does not appear in a graphene, rather a potential step is formed across the interface. The bound states inside this well for $q<0$ can be readily obtained as \cite{Morse1}
  \begin{equation}
  E_{n}^2(q)=(\delta'-\alpha' q^2)^2+\frac{v_F^2}{L^2}\left[q^2-\frac{1}{4}(|q|-2n)^2-\frac{q^4}{(|q|-2n)^2}\right].
  \end{equation}
 Interestingly,  here for any values of $n$, the interfacial modes are sensitive to the width of the interface $L$ which indicates that these modes are purely arising due to the smooth boundary. These states can be also identified as VP states which arise across the interface between irradiated and non-irradiated SD material.\\

\emph{Transport signatures}.
%-------------------------------------------
In this section, we briefly discuss the electron transmission through the interface. First we discuss the transmission probability of an incoming electron with positive energy but with $q>0$ over the interfacial Pöschl-Teller potential well. The transmission probability for an incoming electron with energy ($>0$) across the interface can be obtained as \cite{landau2013quantum}
 \begin{equation}\label{RT}
 T(k_x,q)=\frac{\sinh^2(\pi k_x L)}{\sinh^2(\pi k_x L)+\cos^2[\frac{\pi}{2}(2q+1)]}.
 \end{equation}
 It is interesting to note that an incident electron can pass over the well with $T=1$ without any reflection when $q=j$ with $j=1,2,3..$. This is well-known {\it Ramsauer-Townsedn (RT)} effect. As the depth of the well is determined by $q(q+1)$, the unit transmission can be also regarded as the consequence of the certain depths of the quantum well.  Note that $j=0$ corresponds to the vanishing of the quantum well, hence it is excluded. We note that those $q$ values ($q=j$), for which the unit transmission occurs, correspond to the maxima of the  bound states with respect to the VP index $n$. Hence, these sets of $q$'s actually refer to different VP states which can be also seen from $dE_n/dn=0$. Similarly, for the incident electron with $q<-1$, the transmission probability $T$ can be found by just replacing $(2q+1)$ by $(2|q|-1)$ in the Eq.~(\ref{RT}). Here, $T=1$ at $q=j+1$ and these also correspond to the maxima of bound states with respect to $n$. So it can be seen that the incident electron can fully pass over the well for certain values of $q$ (or the depth of the well $q(q+1)$ or $|q|(|q|-1)$) for which $q$ coincides with the largest VP states. We recall here that in {\it RT} effect over a $1$D rectangular potential well of depth $-V_0$ and width $L$, the unit transmission occurs when the energy of the incident electron ($E_n$) satisfies $E_n+V_0=n^2\pi^2/2\pi L^2$. In contrast, the bound states here are dispersive along the free direction, and only a sets of $q$'s allow unit transmission.
 On the other hand, for the barrier when $-1<q<0$, the transmission probability is 
  \begin{equation}
 T(k_x,q)=\frac{\sinh^2(\pi k_x L)}{\sinh^2(\pi k_x L)+\cosh^2[\frac{\pi}{2}(2|q|+1)]}.
 \end{equation}
which is always less than unity as $\cosh(..)\ge 1$. We shall mention here that the {\it RT} effect is also expected in the irradiated graphene except the term $q$ would become independent of momentum and purely determined by the light parameters. 
 
Finally we comment on the experimental feasibility of realizing VP states. The typical photon energy that is generally used to perturb a quantum system is $0.25$ eV with $eA_0=0.01-0.2\AA^{-1}$. The light parameters belonging to this regime can be used in engineering the VP states.  The recent reported experiment \cite{PhysRevB.96.195104} on identifying the VP states is particularly designed for VP states resulted from the inverted mass term. In the present case also, changing the sign of velocity term across the smooth interface actually changes the sign of the momentum dependent mass term. Hence, similar experimental set-up might be used to detect the VP states in semi-Dirac material. In this experiment, unconventional signature in conductance, charge metastaibility and Hall resistance were identified by a high frequency (RF) measurement of compressibility in high mobility strained $3$D topological insulator using a RF capacitor geometry, which points towards the existence of the VP states.

We shall also comment here on the issue of heating up the driven system. Note that irradiation can heat up the system quickly and the VP states might be affected. In order to keep these states unaffected by the heating, one needs to keep the system in thermally stable. There are several ways of protecting the system from being heated up. One of the most common and efficient approach is the ``Floquet prethermalization'' that helps to prevent the heating of the system. More details about this technique can be found in the recent review Ref.~[\onlinecite{HO2023169297}]
\emph{Summary}.
%-------------------------------------------
We have presented a theoretical study on the VP states which arise across the smooth interface between the two regions of a SD type material which are illuminated by the left and right circularly polarized light, respectively. We have obtained the exact analytical expressions of these modes. These VP states are gapless metallic in nature and strongly sensitive to the momentum. Moreover, we also reveal that the VP states can also exist across the interface between the irradiated or non-irradiated regions of SD material under certain parameter regime. This is also in complete contrast to usual 2D Dirac material like graphene where an interface between irradiated and non-irradiated regions will always act as a potential step instead of well.

%-------------------------------------------
\emph{Acknowledgements}.
%-------------------------------------------
Author acknowledges Tarun Kanti Ghosh and Tutul Biswas for useful comments.

\bibliography{volkov}
%%%%%%%%%% Merge with supplemental materials %%%%%%%%%%

 \end{document}